\documentclass[journal,12pt,onecolumn,draftclsnofoot,letter]{IEEEtran}
\usepackage{tabularx}
\usepackage{amssymb}
\usepackage{graphicx}
\usepackage{subfigure}
\usepackage{fixltx2e}
\usepackage{url}
\usepackage{algpseudocode}
\usepackage{algorithm}
\usepackage{amsmath}
\usepackage{balance}
\usepackage{siunitx}
\usepackage{float}
\usepackage[section]{placeins}
\usepackage{comment}
\usepackage{cite}
\usepackage{tikz}
\usepackage{pgfplots}
\usetikzlibrary{patterns}
\usetikzlibrary{shapes.geometric,positioning,shapes,arrows,calc}
\graphicspath{ {./images/} }
\begin{document}
\title{Local Bitcoin Network Simulator for Performance Evaluation using Lightweight Virtualization}
\author{\IEEEauthorblockN{Lina Alsahan, Noureddine Lasla,  Mohamed Abdallah}

\IEEEauthorblockA{Division of Information and Computing Technology, College of Science and Engineering, HBKU, Doha, Qatar \\ 
Email: \{lalsahan, nlasla, mabdallah\}@hbku.edu.qa}
}

\maketitle
\begin{abstract}
This paper presents a new blockchain network simulator that uses bitcoin's original reference implementation as its main application. The proposed simulator leverages the use of lightweight virtualization technology to build a fine tuned local testing network. To enable fast simulation of a large scale network without disabling mining service, the simulator can adjust the bitcoin mining difficulty  level to below the default minimum value. In order to assess the performance of  blockchain under different network conditions, the simulator allows to define different network  topologies, and integrates Linux kernel traffic control (tc) tool to apply distinct delay or packet loss on the network nodes. Moreover, to validate the efficiency of our simulator we conduct a set of experiments and study the impact of the computation power and network delay on the network's consistency in terms of number of forks and mining revenues. The impact of applying different mining difficulty levels is also studied and the block time as well as fork occurrences are evaluated. Furthermore, a comprehensive survey and taxonomy of existing blockchain simulators are provided along with a discussion justifying the need of new simulator. As part of our contribution, we have made the simulator available on Github\footnote{ \url{ https://github.com/noureddinel/core-bitcoin-net-simulator}} for the community to use and improve it.

\end{abstract}

% The invention of blockchain technology has radically changed the design operation of distributed systems where it has been considered as a potential solution for many global problems which concerns nations' most critical sectors such as energy, finance and health. Amongst the available blockchains with the varying architectures and consensus mechanisms, Bitcoin is considered as the bedrock of this field since its introduction in 2008, where it received most of the attention and gained a value reached \$41 billion. The underlying protocol of Bitcoin is expected to be the future's trusted peer-to-peer transaction ecosystem. However, Bitcoin suffers from scalability limitations linked to the public network constraints. 
% %Moreover it allows only a limited number of transactions to be processed at a period of time compared to other transactional systems which hinders the usability of Bitcoin since each transaction requires on average 10 minutes to be verified.
% Bitcoin-ng \cite{NG}, GHOST \cite{GHOST} and other platforms \cite{messagePipelining} were proposed to address Bitcoin's limitations by modifying its architecture and work flow. At the same time, ongoing academic researches are studying the original bitcoin protocol performances and security with the aim of understanding the protocol's bottlenecks and improving its performances \cite{SimBlockEvent}. 

\section{Introduction}

Blockchain is one of the most promising technologies of the last decade, with its potential of solving several issues in the current distributed and peer-to-peer ecosystems. Blockchain provides a secure data sharing/processing/storing, and a trusted environment for untrusted communicating systems without relying on any kind of intermediaries \cite{zheng2017overview}. Although potentially blockchain brings key security benefits, traditional blockchain platforms are still suffering from serious scalability issues to meet high transactional throughput, low  latency and other issues related to computation and storage usage. For instance, the first known cryptocurrency, Bitcoin, has a transaction throughput of only 7 transactions per second (TPS) \cite{bitcoin_chart}, while Ethereum can achieve about 14 TPS \cite{ethereum}, which is still not enough to be adopted in systems with high frequency
transactions requirement. To overcome these limitations, several research groups are working in an attempt to study \cite{blockbench, SimBlockEvent} and improve \cite{sharding1,omniledger, Rapidchain} the performance of blockchain and to come up with a balanced architecture that can be widely adopted in practice. 

One of the necessary tools to assist performance evaluation and understand blockchain's behavior under a close to realistic environment conditions are simulators. In the literature, the existing blockchain performance evaluation approaches can be classified into three categories according to their architecture and run time operations. First, \textit{event-based simulator} \cite{VIBESEvent,SimBlockEvent,shadoworiginalclient&eventmining}, that simulate blockchain network by abstracting the node logic or part of it into a set of discrete events triggered at an instant of time based on the original run time of a blockchain node. Although an event-based simulator is a scalable and  cost-effective approach, it might ignore some minor, yet important traits of blockchain due to the abstraction of the nodes' functionalities. Moreover, developing an event-based simulator is a complex task and time consuming since the blockchain client application needs to be re-implemented. Second, \textit{historical-data analysis} is another method of evaluating blockchain network. Logs are collected from monitoring nodes \cite{heoretical} which are connected to the main  network, and  then analysed to refine the results. This approach suffers from multiple limitations such as the collected data are under the actual blockchain network factors and can not be parameterized to study the blockchain under other network conditions. Moreover, the collected data is sourced from the logs of a limited number of nodes and does not provide the ability of observing the rest of the network nodes. Last, \textit{virtualization-based simulator} \cite{virtualization}, where virtual multi-nodes blockchain network is constructed using a set of containers that runs blockchain reference client and manages the interconnections between the nodes. Lightweight virtualization technique allows such frameworks to be portable and easy to setup while maintaining the reference client application unchanged which adds accuracy to the conducted experiments and results analysis. \cite{virtualization} is one framework that follows this approach, however, it lacks a configuration layer that assists researchers to setup specific testing scenarios to study multiple metrics of interest.   

\tablename~\ref{table:comparison} provides a high level comparison between the different discussed simulation categorise based on architectural, operational and performance aspects. From the table we can conclude that virtualization-based frameworks provides relatively high usability, configuration and accuracy compared to event-based frameworks and historical data analysis. These metrics are considered as key features for any network simulators, specially when assessing the performance of financial applications such as Bitcoin. For the complexity, speed and scalability, still virtualization-based approach provide acceptable moderate performance that can be improved by leveraging high computation power machine.

\begin{table}[h!]
\centering
\caption{Blockchain network analysis frameworks comparison}
\begin{tabular}{ |p{2.3cm}||p{1.7cm}|p{1.7cm}|p{1.5cm}|}
 \hline
Comparison aspects & Discrete-event & Virtualization-based  & Historical data\\
 \hline
 Complexity & high & moderate & low\\
 \hline
 Usability & moderate & high & moderate\\
 \hline
 Configuration	&  moderate & high & low\\
 \hline
 Accuracy &	low	& high & high\\
 \hline
 Speed	& high	& moderate &	low\\
 \hline
 Scalability &	high & moderate & low\\
 \hline
 \end{tabular}
 \label{table:comparison}
 \end{table}

In this paper, we propose a new blockchain network simulation framework for Bitcoin performance analysis and evaluation. We adopt the virtualization approach for constructing a private network through executing Bitcoin's full node reference application with the ability of performing fine tuned testing scenarios. Our simulation framework integrate network emulation tool to assist the impact of different practical network constraints. Moreover, the framework automatically tunes the mining difficulty level of Bitcoin to enable fast simulation by considering the actual available computation resources and the desired size of the network. This feature is beneficial when the host machine used in the experiment is not capable of performing intensive computations. Alternatively, the nodes can be also configured to mine  following the actual difficulty adjustment implemented in the official Bitcoin client. In addition, the simulator allows to manually allocate the available mining resources between the network nodes to fulfill the experiments purposes.
We prove the efficiency and usability of our simulator by conducting a set of experiments to evaluate Bitcoin network performance under different network and mining conditions. The impact of different mining difficulties level on the network stability and throughput is also studied. 

The rest of the paper is organized as follows. In Section \ref{sec:framework}, we describe the framework architecture and its functionalities. In Section \ref{sec:evaluation}, we present the network setup and the conducted experiments, as well as discussing the obtained results. Finally, in Section \ref{sec:conclusion}, we conclude our work and discuss
future improvements. 

\section{Simulation Framework Architecture}
\label{sec:framework}
\begin{figure}[h]
\centering
\includegraphics[width=9cm, height=6.5cm]{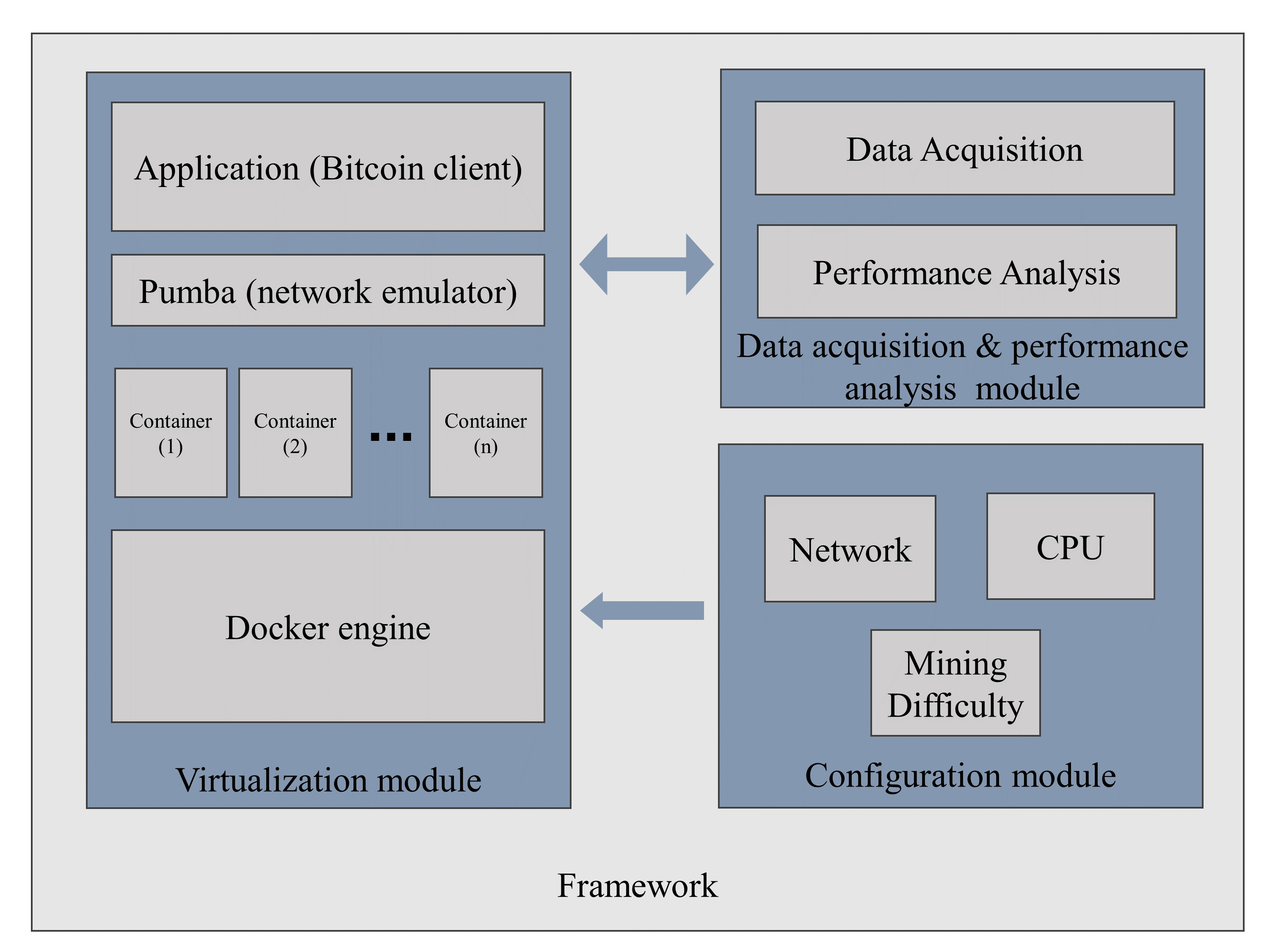}
\caption{Framework architecture}
\label{fig:framework architecture}
\end{figure}

In this section, we present the architecture of our virtualization-based simulation framework.
As shown in \figurename~\ref{fig:framework architecture}, the simulator is composed of three major modules, Virtualization module, Configuration module and Data acquisition and Performance analysis module. Each consists of multiple units which carry out vital tasks in run time. In the following sections, we give a detailed description and the role of each module in achieving the goal of the simulator.

\subsection{Virtualization module}
\subsubsection{Virtual hosts and network nodes}
Blockchain nodes are operating in virtual hosts by utilizing docker containers which provides portability, isolation and lightweight virtualization \cite{docker}. Each container runs the Bitcoin core reference client application.
%version 0.1.13 \cite{}. This version is selected as it is the last version allowing to perform mining using CPU, newest versions require an external ASICs mining hardware.
In case of multiple host machines are available for simulation, docker swarm
can be utilized to create an overlay network between the different physical machines. Thus, evaluating a very large blockchain network is possible using our framework in the docker swarm mode.

\subsubsection{Network emulation}
To emulate real life network conditions, our framework uses Pumba chaos testing tool\cite{pumba}, which was developed to evaluate docker containers application resiliency in run time and ensure the reliability of the system and its ability to recover from failures. Pumba is also featured with an option to enable network testing, by emulating wide variety of realistic network properties, such as network delays, bandwidth and packet loss.
Pumba manipulates containers' egress traffic at the kernel level, by using the built in capabilities of Linux that allows altering the traffic control (tc) rules through the netem extension. Pumba allows user to specify which network action to take for which period of time, container, network interface, link and direction (in-bound/out-bound).  

\subsubsection{Bitcoin reference application}
Bitcoin core provides alternative blockchains for testing and development purposes. These testing environments suffer from limitations and drawbacks which may hamper their utilization for testing and evaluating eventual solutions. For instance, Testnet  is dedicated for developers to explore Bitcoin network without risking in dealing with real cryptocurrency. Testnet is designed to only mimic the actual Bitcoin network and does  not allow  to perform any modification on the client's application source code. Another alternative is Regtest network which was developed for the same purpose yet, it does not include the mining process, instead, blocks are generated instantly on demand.  

In our framework we have altered the Testnet Bitcoin network client application to build a local network disconnected from the public Testnet network, which provides full control over the testing environment. To obtain a new blockchain we have generated a new genesis block using a python script that returns the merkle hash and genesis hash, and changed the original genesis information in the source code. In addition DNS seeds were removed to prevent the nodes from continuously attempting to connect to the hard coded peer IP addresses. In our framework we have used Bitcoin client version 0.10.2\footnote{\url{https://bitcoin.org/bin/bitcoin-core-0.10.2/}}. This version is selected as it is the last one allowing internal mining using CPU, newest versions require an external ASICs mining hardware. 

\subsection{Configuration module}
Our Framework is highly configurable and allows tuning the evaluation environment to assist accurate observations. We will discuss three major configurations that concern the network condition, mining power and mining difficulty. Our framework can be further upgraded to support other possible configurations without major changes.
\subsubsection{Network configuration}
The underlying topology and its network metrics play crucial role on deciding the performance of blockchain. To provide high network configuration flexibility,  our framework allows to customize the blockchain topology by specifying the number of nodes as well as the connectivity between them. In addition, a customized amount of delay on specific nodes or links can be also introduced with a given start and end time. 

\subsubsection{Mining configuration}
 Our framework provides discrete distribution of the host machine resources among the running nodes. According to the available CPU power on the host machine(s), users can specify the number of CPUs to assign to each individual node. Mining nodes can be configured by enabling the internal miner of the Bitcoin reference client to continuously generate new blocks and by utilizing the maximum resources assigned to each node based on the user's choice. This is an important feature that allow to study how the allocation of power resources affect the security of the whole network as well as the revenue of each individual node.
 
\subsubsection{Mining difficulty configuration}
Mining difficulty is the pivot of a bitcoin network, it controls its performance and its primary parameters. In a restricted testing environment with limited computation power, mining difficulty must be well configured to fit the testing network topology. The official Bitcoin-core implementation set the minimum difficulty to $1$, which is still expensive in term of computation and can lead to high block time, especially in the case of our simulator that share the resources of the host machine among all the simulated nodes. For this reason,  we propose to automatically adjust the mining difficulty $C_{\text{diff}}$ and go below the default value, by considering  the configured number of nodes $M$, target block time $t$, and available computation power (hash-rate) $H$ of each node, using the following  formula (\ref{eq:1}).

\begin{equation}\label{eq:1}
C_{\text{diff}} = \dfrac{t \sum_{i=1}^{M}(H_i)}{2^{32}} % 2^32 diff relatively to the max diff
\end{equation}
The impact of applying different difficulty levels on the  consistency of the network in terms of number of forks and block time will be discussed  in Section \ref{sec:evaluation}. 

\subsection{Data acquisition and performance analysis module}
\begin{figure}[h]
\centering
\includegraphics[width=9cm, height=7cm]{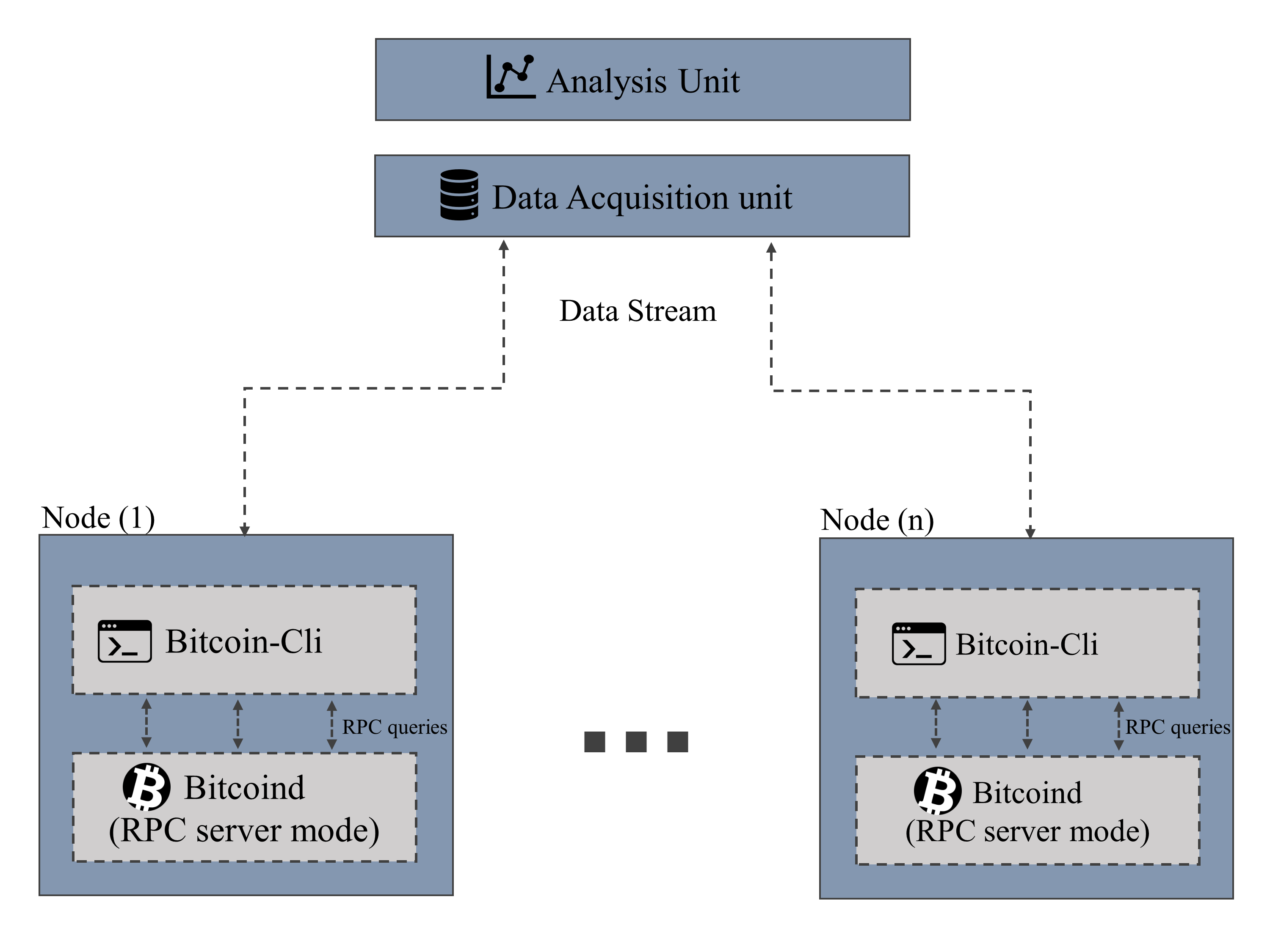}
\caption{Data acquisition and performance analysis module}
\label{fig:Data process}
\end{figure}

As shown in \figurename~\ref{fig:Data process}, the data acquisition and performance analysis module is responsible of collecting real time data from the running nodes and process them in runtime to generate performance reports during and at the end of the simulation. Blockchain state values are gathered through RPC quires to the bitcoind servers which is Bitcoin's node application running in RPC server mode. 

\subsubsection{Data acquisition unit}
The data logging unit issues request of streaming recent data, using Bitcoin client  command line interface which in turn sends RPC quires to the bitcoind server to retrieve a set of the requested values. Namely, the (i) mining state, (ii) number of generated blocks and (iii) number of forks. Retrieved data is then stored in log files or processed instantly per request for generating performance reports. 
\subsubsection{Performance analysis unit}
This unit is adaptable to the needs of the simulation by including various type of calculations and operations to be performed on the logged data. Currently, the framework by default calculates the following; (1) Number of generated blocks by each miner, (2) number of blocks committed to the main chain by each mining node, (3) number of valid forks each node has witnessed, (4) amount of spendable balance of each node, (5) current hashrate of the network and (6) nodes with the highest and lowest number of mined blocks. Plots are also automatically generated with the report to assist the understanding of the obtained results.  

\section{Bitcoin Network Evaluation}
\label{sec:evaluation}
%nderstanding and Quantifying the constraints and limitations of Bitcoin blockchain \figurename~\ref{fig:evaluation} has been given an immense attention in the research community to provide a clear guide for adopting the technology in industrial use cases

\begin{figure}[]
\centering
\scalebox{.70}{
\begin{tikzpicture}
  \node (core) at (0,0) [draw,very thick,dashed, red, minimum width=8.3cm,minimum height=2cm] {Evaluation Framework};
         %----------Metrics------------------------------------
       \node [above left=0.2cm and -2.05cm of core](tps) [draw,thick,minimum width=2cm,minimum height=1cm] {TPS};
        \node [ right=0.05cm of tps](latency) [draw,thick,minimum width=2cm,minimum height=1cm] {Latency};
        \node [ right=0.05cm of latency](fork) [draw,thick,minimum width=2cm,minimum height=1cm] {Nb. Forks};
        \node [ right=0.05cm of fork](reward) [draw,thick,minimum width=2cm,minimum height=1cm] {Reward};
        \node [left=0.2cm  of tps](text3)  {Metrics};
        %------------Net. Constraints-------------------------  
        \node [below left=0.2cm and -2.05cm of core](delay) [draw,thick,minimum width=2cm,minimum height=1cm] {Delay};
       \node [ right=0.05cm of delay](topology) [draw,thick,minimum width=2cm,minimum height=1cm] {Topology};
        \node [ right=0.05cm of topology](link) [draw,thick,minimum width=2cm,minimum height=1cm] {Link Quality};
       \node [ right=0.05cm of link](size) [draw,thick,minimum width=2cm,minimum height=1cm] {Net.  Size};
       \node [left=0.2cm  of delay](text2)  {Net. Constraints};
       %-------------Evaluation----------------------------
        \node [above left=0.2cm and -4.1cm of tps](eff) [draw,thick,minimum width=4.05cm,minimum height=1cm] {Efficiency};
        \node [above left=0.2cm and -4.1cm of fork](sec) [draw,thick,minimum width=4.05cm,minimum height=1cm] {Security / Fairness};
        \node [left=0.2cm  of eff](text4)  {Performances};
       %-------------Sys. const.---------------------------
        \node [below left=0.2cm and -4.12cm  of delay](sys) [draw,thick,minimum width=4.1cm,minimum height=1cm] {Computation Power};
        \node [right=0.06cm  of sys](sys2) [draw,thick,minimum width=4.1cm,minimum height=1cm] { Memory Space};
        \node [left=0.2cm  of sys](text1)  {Sys. Constraints};
        %-------------App. const.---------------------------
        \node [below left=0.2cm and -4.13cm of sys](const) [draw,thick,minimum width=4.1cm,minimum height=1cm] {Sending Rate};
        \node [ right=0.06cm  of const](const2) [draw,thick,minimum width=4.1cm,minimum height=1cm] { Number of Senders};
        \node [left=0.2cm  of const](text1)  {App. Constraints};
\end{tikzpicture}
}
\caption{System evaluation framework}
\label{fig:evaluation}
\end{figure}
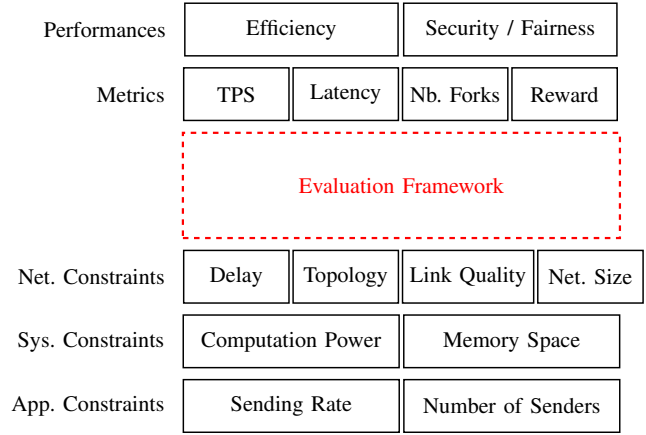

In this section, we evaluate the performance of Bitcoin network using our proposed simulator. We summarise in \figurename~\ref{fig:evaluation} the different metrics and constraints that should be evaluated and considered by the simulator, respectively.  For Bitcoin network, four performance metrics, namely the transactions throughput (TPS), network latency, number of forks and mining rewards can me evaluated to assess the efficiency,  security and fairness of the system. For the parameters that may affect the Bitcoin network,  we distinguish between network, system and application constrains. Network topology and the quality of the links between peers have inevitable impact on the network performance, where as the computation power and memory space dedicated by each mining node are considered as system constraints and directly affect the miner rewards. At the application level, the number of senders or the sending rate can also affect the efficiency of the system.

 Most of the existing studies that evaluated the performance of Bitcoin network have focused on the impact of block propagation delay and its relation with the produced number of forks and its concomitant impact on both security and usability of the network \cite{delay1, delay2, delay3, delay4}. In this paper, in addition to studying the network delay caused by the peer-to-peer communication, we  introduce an empirical analysis to assess the effect of available mining power on the miners' revenues (rewards), as there are not many studies in this context. Also we evaluate the impact of different mining difficulty levels on the network consistency and throughput.  

\begin{figure}[h]
    \centering
    \includegraphics[scale=0.3]{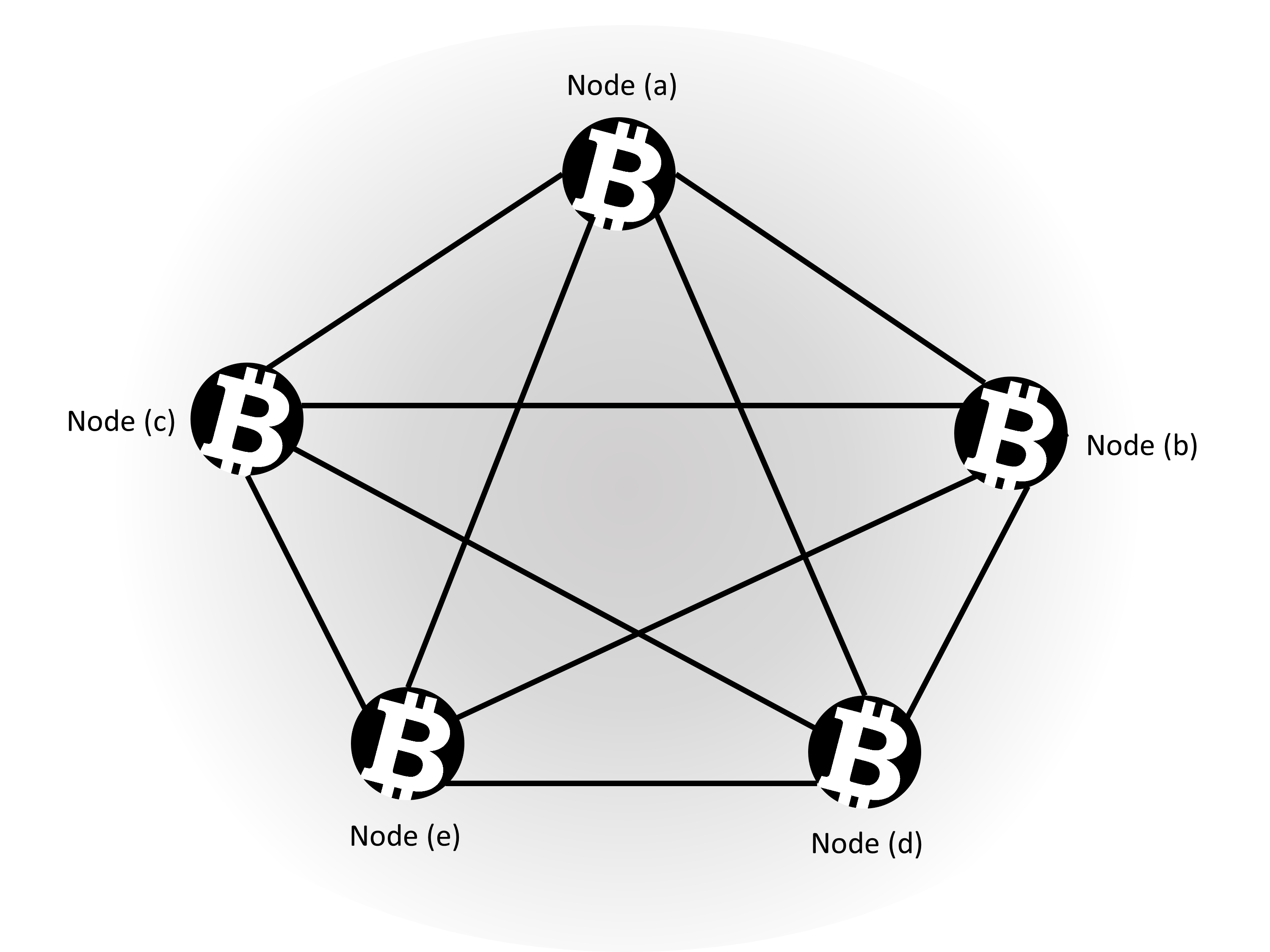}
    \caption{Local Bitcoin test network with mesh topology}
    \label{fig:topology}
\end{figure}

We have setup a testing network that consists of five full nodes where each node is connected to all other online nodes forming a mesh topology, as shown in \figurename~\ref{fig:topology}. The experiments are conducted on a workstation machine with Intel(R) Xeon(R) Gold 6130 CPU, 2.10 GHz, 64 core CPU, 256GB RAM, and running Ubuntu 18.04.2.

\subsection{Mining Resources}
Bitcoin mining is a resource intensive process and requires clear understanding of the relationship between the cost of mining and the corresponding obtained rewards. In the experiment, the miner's revenue is calculated as the miner's committed blocks multiplied by the block's reward. The hash power used during mining can be represented in terms of hash rate percentage, and is calculated using formula (\ref{eq:2}). In the formula, $H\%$ represents the percentage of hashing power that is owned by miner $i$ with respect to the hashing power of the entire network, and $M$ is the total number of miners. 

\begin{equation}\label{eq:2}
H_i\% = \dfrac{H_i } {\sum_{j=1}^{M}(H_j)}
\end{equation}

In the experiment, the mining difficulty is fixed to 1, the simulation duration is set to 24 hours, and the following number of CPUs (mining power) 4, 8, 12, 16 and 20 were attributed to Node(a), Node(b), Node(c), Node(d) and Node(e), respectively. We report in  \figurename~\ref{fig:hashplot} the number of committed block for each mining node. The plot clearly shows that the number of committed blocks to the main chain increases almost linearly with the mining power of the node. Which yields a linear increment in the mining rewards (BTC). However increasing the hashrate percentage in the range from 7\% to 20\% does not yield a significant benefit as in the range from 20\% to 33\% although the amount of increment in both ranges equals 13\%. Miners tend to adjust their share in the total network hashing power by investing in more mining infrastructure, however, estimating the benefit that will be gained from increasing the mining hashrate is paramount to avoid power waste with scant rewards.

\usetikzlibrary{datavisualization}
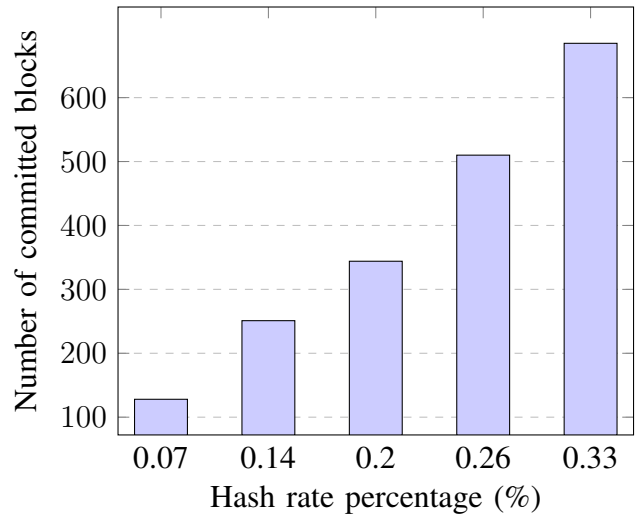
\begin{figure}
\centering
\begin{tikzpicture}
\usetikzlibrary{patterns}
\begin{axis}[
	x tick label style={symbolic x coords={0.07,0.14,0.2,0.26,0.33},
	/pgf/number format/1000 sep=},
	ylabel= Number of committed blocks, xlabel=Hash rate percentage (\%),
	enlargelimits=0.1,
	legend style={ anchor=south,legend columns=1}, xticklabel style = {xshift=-0.0cm},
	 ymajorgrids=true, grid style = dashed, ymax=686, bar width=20pt,
	  ytick={100,200,300,400,500,600}, legend pos=north west]
\addplot 
	[ybar,fill=blue!20]  %216    %341
	coordinates { (0.07, 128) }; 
\addplot  %123
	[ybar, fill=blue!20] 
	coordinates  { (0.14, 251) };
\addplot %93
	[ybar, fill=blue!20] 
	coordinates  { (0.2, 344) };
\addplot  %166
	[ybar, fill=blue!20] 
	coordinates { (0.26, 510) };
\addplot  %175
	[ybar, fill=blue!20] 
	coordinates { (0.33, 685) };
% xmax=5,
%xtick={1,2,3,4,5,6},
% \legend{7\%,
% 14\%,
% 20\%,
% 26\%,
% 33\%}
\end{axis}
\end{tikzpicture}
\caption{Impact of hashing-rate share on miners' committed blocks (income)}
\label{fig:hashplot}
\end{figure}

\subsection{Network Delay}
Bitcoin nodes use a gossiping protocol to propagate blockchain announcements and data through a peer-to-peer network. This protocol is susceptible to network delays that is the first cause of blockchain forks. Although forks are considered as a normal behavior of Bitcoin network, it indicates inconsistency in the network and can hinder its security. Moreover network delays may influence the network fairness, because nodes that experiences higher propagation delay will not have the chance of committing their generated blocks to the main chain before other competing blocks. To assess the relation between network delays, network fairness and number of forks, we have conducted an experiment that applies different network delays ranging from 15 seconds to 1 minute on the egress traffic of each node's container. The obtained results from the experiment are depicted in \figurename~\ref{fig:blocksdelay} and \figurename~\ref{fig:forksdelay}. We can clearly notice that the number of committed blocks to the main chain decreases exponentially with the increase of network delay. Thus, mining nodes should also consider investing in high link quality similar to  mining power to achieve better rewards. The result obtained from \figurename~\ref{fig:forksdelay} proves our former discussion where the frequency of fork occurrence manifest an incremental relation with the network delay.    

\usetikzlibrary{datavisualization}
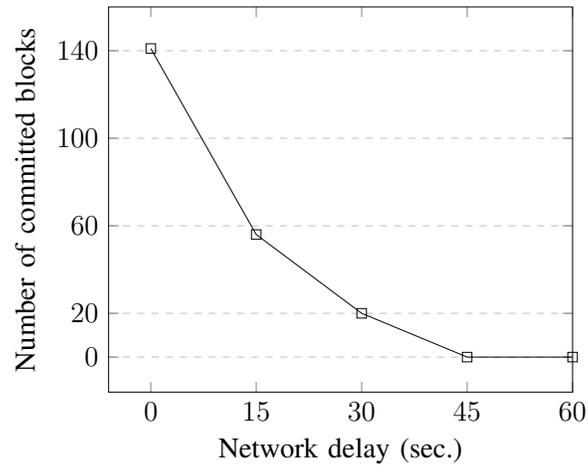
\begin{figure}
    \centering
    \scalebox{0.9}{
    \begin{tikzpicture}
        \begin{axis}[xmax=60,ymax=160,samples=10,grid=minor,xlabel={Network delay (sec.)},ylabel={Number of committed blocks},legend style={ anchor=south,legend columns=1}, xtick={0,15,30,45,60}, ytick={0,20,60,100,140},ymajorgrids=true,
    grid style=dashed]
            \addplot[
    color=black,
    mark=square,
    ] coordinates {
    
            (0,141)
            (15,56)
            (30,20)
            (45,0)
            (60,0)

            };

        \end{axis}
    \end{tikzpicture} }
\caption{Impact of network delays on the number of committed blocks to the main chain}
\label{fig:blocksdelay}
\end{figure}

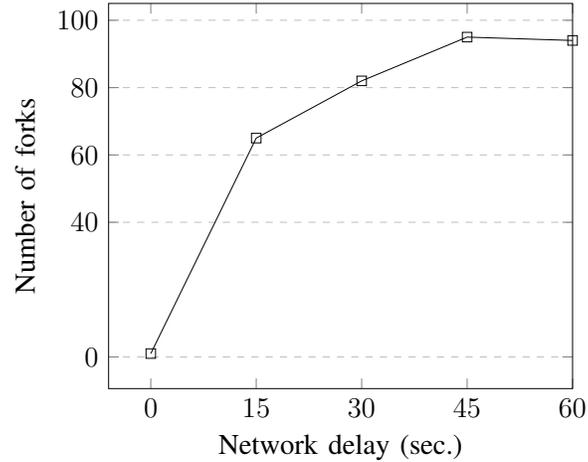
\begin{figure}
  \centering
    \scalebox{0.9}{
\begin{tikzpicture}
\usetikzlibrary{patterns}
\begin{axis}
[xmax=60,ymax=105,samples=10,grid=minor,xlabel={Network delay (sec.)},ylabel={ Number of forks}, xtick={0,15,30,45,60}, ytick={0,40,60,80,100},ymajorgrids=true,
    grid style=dashed]

\addplot [
    color=black,
    mark=square,
    ]coordinates { (0,1) (15,65) (30,82) (45,95) (60,94)}; 
	
\end{axis}
\end{tikzpicture}}
\caption{Impact of network delays on fork occurrences}
\label{fig:forksdelay}
\end{figure}

\subsection{Mining Difficulty}

%add one introduction sentence 
In this section, the impact of applying different mining difficultly level on the block time and forks occurrence is studied. This study will help in finding the suitable difficulty level that can be defined to build a stable testing network, that balance between the simulation's efficiency, and the network's consistency and security. In the experiment, the average Block time and number of forks are evaluated under different mining difficulty levels, ranging from 0.001 to 0.1. The testing network is configured with five mining nodes each granted equal mining resources which is set to 10 CPUs. In the experiment, we mean by mining difficulty level the difficulty to reach the target (nBits) of the hashing puzzle, and it is determined using formula (\ref{eq:3}), where $Max_{t}$ corresponds to the maximum possible target.

\begin{equation}\label{eq:3}
Diff = \dfrac{Max_{t}}{nBits} 
\end{equation}

\usetikzlibrary{patterns}
\usetikzlibrary{datavisualization}
\begin{figure}
\centering
\begin{tikzpicture}
\begin{axis}[
	x tick label style={
	symbolic x coords={0.001,0.005,0.01,0.05,0.1, 0.2},
	/pgf/number format/1000 sep=},
	enlargelimits=0.01,ylabel= Block time (sec), xlabel= Mining difficulty,
	ymajorgrids=true, xmajorgrids=false, grid style = dashed, ytick={0,1,5,10},
	legend style={at={(0.5,-0.1)},
	anchor=north,legend columns=-1},
	ybar interval=0.7, xmajorgrids=false, 
]
\addplot [fill = blue!20]%block time
coordinates {(0.001,0.1109) (0.005,0.5429)
		 (0.01,1.3250) (0.05,6.679)(0.1,11.2418) (0.2,11.2418) };
		 
\end{axis}
\end{tikzpicture}
% \label{fig:hashplot}
 \caption{Impact of mining difficulty level on the average block time}
\label{fig:diffinterblock}
\end{figure}
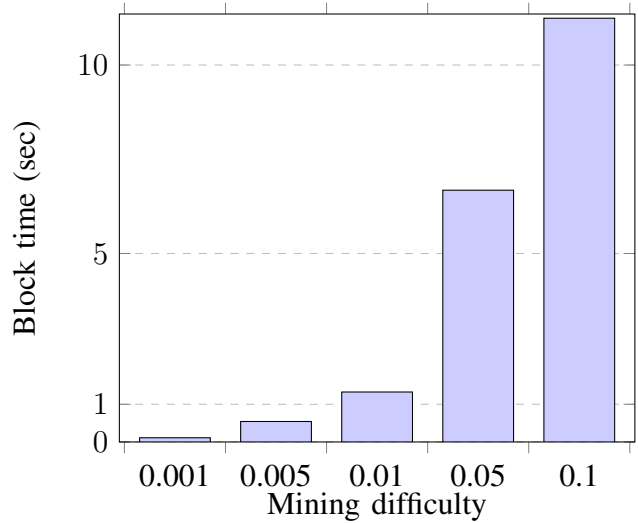

\begin{figure}
\centering
\begin{tikzpicture}
\begin{axis}[
	x tick label style={
	symbolic x coords={0.0001,0.0005,0.001,0.005,0.01,0.05,0.1,0.2},ylabel= Number of forks, xlabel= Mining difficulty,
	/pgf/number format/1000 sep=},ytick={20,50,100,200},
	enlargelimits=0.02,
	legend style={at={(0.5,-0.1)},
	anchor=north,legend columns=-1},
	ybar interval=0.7, ymajorgrids=true, xmajorgrids=false, grid style = dashed
]

\addplot [ fill = red!20 ]%forks
	coordinates {(0.001,300) (0.005,178)
		 (0.01,103) (0.05,26.6)(0.1,13)(0.2,13)};
\end{axis}
\end{tikzpicture}
\caption{Impact of mining difficulty level on the average number of forks}
 \label{fig:diffforks}
 \end{figure}
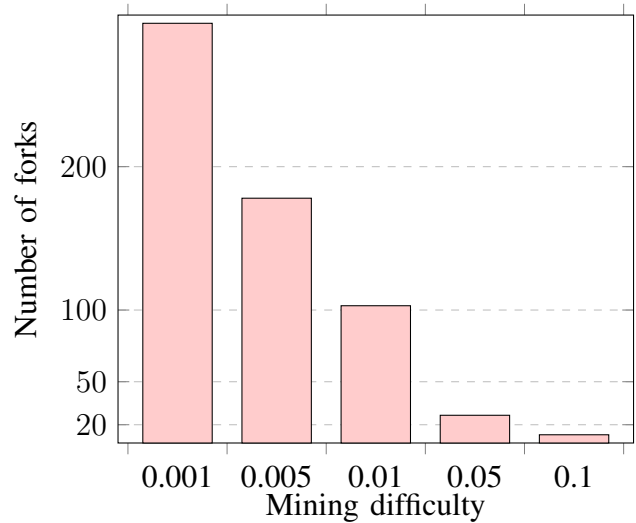

The experiment's results are depicted in \figurename~\ref{fig:diffinterblock} and shows that under low mining difficulty level, the network has noticed a higher blocks throughput since the block time is scant and close to instant. However, the average number of  occurred forks, as shown in  \figurename~\ref{fig:diffforks}, is inversely  proportionally to the mining difficulty level. The fast pace of generating new blocks by the competing miners increases the probability of having multiple nodes that simultaneously find new blocks of the same height, and eventually leads to blockchain fork. Tuning the simulator to mine with a difficulty of 0.05 allows setting up an efficient testing network while avoiding disturbing the network's consistency and enables generating the first 2016 blocks in in 3 hours which assist in performing fast experiments in stable and efficient testing environment.

\section{Conclusion}
\label{sec:conclusion}
In this paper, we have proposed a new blockchain network  simulation framework that is based on a lightweight virtualization technology, and is highly configurable. In order to mimic actual network conditions, the simulator  allows to introduce several network constraints, such as network topology and delay, assign different mining powers, and automatically adjust the mining difficulty level to guarantee reasonable simulation time.  
To assess the  efficiency of the developed  simulator, we have conducted several experiments  and studied the impact of network delays, hashing power and mining difficulty on the consistency and fairness of the Bitcoin network.  
Further experiments can be conducted using our simulator to evaluate Bitcoin network under other network scenarios. Moreover, the modular architecture of the simulator make it easier to introduce new features, such as supporting configurable security attacks and point of failure resiliency tests, without making major changes to the simulation framework.

\bibliographystyle{IEEEtran}
\bibliography{References}

\end{document}